\documentclass[journal]{IEEEtran}

\IEEEoverridecommandlockouts

\usepackage{algorithm}
\usepackage{algorithmic}
\usepackage{balance}
\usepackage{subcaption} 
\usepackage{caption}
\usepackage{float}
\usepackage{enumitem}
\usepackage{booktabs}
\usepackage{balance}
\usepackage{cite}
\usepackage{amsmath,amssymb,amsfonts}
\usepackage{bbm}
\usepackage{bm}
\usepackage{hyperref}
\usepackage{amsthm}
\usepackage{graphicx}
\usepackage{xcolor}
\usepackage{stfloats}
\usepackage{amsthm}
\usepackage{amsmath}

\makeatletter%
\if@twocolumn%
\else
\fi%
\makeatother%

\allowdisplaybreaks

\def\bG{\mathbf{G}}

\def\bSigma{\boldsymbol{\Sigma}}

\def\btheta{\boldsymbol{\theta}}

\def\trace{\text{tr}}

\DeclareFontFamily{U}{mathx}{\hyphenchar\font45}
\DeclareFontShape{U}{mathx}{m}{n}{
      <5> <6> <7> <8> <9> <10>
      <10.95> <12> <14.4> <17.28> <20.74> <24.88>
      mathx10
      }{}
\DeclareSymbolFont{mathx}{U}{mathx}{m}{n}
\DeclareFontSubstitution{U}{mathx}{m}{n}
\DeclareMathAccent{\widecheck}{0}{mathx}{"71}
\DeclareMathAccent{\wideparen}{0}{mathx}{"75}

\begin{document}


\title{AISAC: Closing the Loop Between AI and Integrated Sensing and Communication for 6G}

 \author{Mehdi~Karbalayghareh,~\IEEEmembership{Member,~IEEE,} Abhishek~Rajasekaran,~\IEEEmembership{Student Member,~IEEE,} Xiaoyan~Ma,~\IEEEmembership{Member,~IEEE,} David~J.~Love,~\IEEEmembership{Fellow,~IEEE,} \\ and~Christopher~G.~Brinton,~\IEEEmembership{Senior Member,~IEEE}

\thanks{M. Karbalayghareh, A. Rajasekaran, X. Ma, D. J. Love, and C. G. Brinton are with the Department of Electrical and Computer Engineering, Purdue University, West Lafayette, IN, USA (e-mails: mkarbala@purdue.edu; rajasek1@purdue.edu; ma946@purdue.edu; djlove@purdue.edu; cgb@purdue.edu).}}

\maketitle

\begin{abstract}
Integrated sensing and communication (ISAC) and AI-and-communication (AIAC) are identified as separate usage scenarios in the ITU IMT-2030 vision for sixth-generation (6G) networks. In practice, however, these two directions are already beginning to merge. ISAC gives the network a way to observe the physical world, while AI gives the network a way to learn from those observations and act on them. This article introduces \emph{AI-integrated sensing and communication (AISAC)} as a closed-loop framework for this merger. In AISAC, AI is not only a tool used to optimize an ISAC system. ISAC is also the physical substrate through which AI receives data, context, and connectivity. The key technical message is that AISAC requires a new physical-layer design principle, in which the ISAC waveform, beam, power, bandwidth, and sensing mode should be configured for \emph{learning alignment}, not for sensing distortion or communication rate alone. In particular, the sensing configuration that is most accurate from a classical estimation viewpoint need not be the one that is most useful for training or inference. We present the AISAC landscape, explain why imperfect sensing changes the learning problem, develop the closed-loop architecture and its three-way sensing--communication--learning tension, and outline a vehicular edge-intelligence use case together with open problems for theory, implementation, and standardization.
\end{abstract}


\section{Introduction}
\label{sec:introduction}
Sixth-generation (6G) wireless networks are expected to move beyond connectivity as the only native service of the radio access network~\cite{Giordani-6G-ComMag2020,Brinton-Key6G-CommMag25}. Two of the clearest signals of this shift are \emph{AI-and-communication (AIAC)} and \emph{integrated sensing and communication (ISAC)}, both identified in the ITU IMT-2030 framework as usage scenarios for networks toward 2030 and beyond~\cite{ITU_M2160}. AIAC reflects the view that future networks will not simply carry artificial intelligence (AI) traffic; they will also support model training, inference, adaptation, and AI-driven control as first-class network functions. ISAC reflects a parallel view: the same infrastructure that delivers data can also sense the environment, track objects, localize users, and build situational awareness from radio signals.

These two directions are often developed separately. AIAC discussions usually emphasize distributed training, edge inference, model transport, and AI-native air-interface control. ISAC discussions usually emphasize shared waveforms, radar-like sensing, localization, beam management, and the tradeoff between sensing accuracy and communication rate. Yet this separation is becoming artificial. A network that senses its environment will naturally generate data for AI. A network that runs AI will naturally use the learned model to decide where, when, and how to sense and communicate. The two capabilities are therefore not independent services placed side by side; they form a feedback loop.

In this article, we introduce \emph{AI-integrated sensing and
communication (AISAC)}, which denotes the closed-loop integration of AI with ISAC. AISAC has two directions. In the first direction, AI improves ISAC by learning the models that can configure beams, waveforms, resource splits, tracking policies, interference management, and network orchestration. In the second direction, ISAC improves AI by providing the sensing and communication layers that can produce the observations, labels, context, localization information, mobility information, and model-exchange channels on which AI depends. The central point is that these two directions should not be treated as two isolated blocks. They should be designed as one loop in which the current AI model changes the next ISAC acquisition policy, and the resulting sensed data and communication outcomes change the next AI model.

Emerging 6G verticals further motivate this closed-loop view. Automotive systems, industrial automation, smart cities, eHealth, public safety, and immersive services require not only connectivity, but also continuous perception of the physical environment and adaptation to task-level objectives~\cite{NextGAllianceVerticals2023}. In these settings, sensed observations are not merely auxiliary measurements; they become inputs to learning systems that support prediction, control, mapping, and decision-making. AISAC is therefore naturally aligned with vertical applications in which sensing quality, communication reliability, and learning utility must be optimized together.

This viewpoint creates a design objective that neither sensing nor communication captures on its own. It is a cross-layer principle. The model's current sensitivity determines how the shared physical layer should acquire and deliver information:

\color{black}
\begin{quote}
\emph{The AISAC physical layer should be configured to acquire and deliver information that is useful for the AI task, while satisfying the sensing-mission and communication requirements of the network.} 
\end{quote}
\color{black}

This principle matters because sensed data are acquired from the physical environment and then become input to learning and control. They act as training samples, inference inputs, radio maps, digital twin updates, or context for control. A sensing mode with a low mean-squared error may still be poor for learning if its errors fall in directions to which the model is highly sensitive. Conversely, a sensing mode with a larger classical distortion may be more useful if its uncertainty lies mostly in directions that do not affect the learning task. AISAC therefore asks a different question from conventional ISAC: not only ``how accurately did the network sense?'' or ``how fast did the data communicate?'', but also ``how useful was the acquired information (through noisy sensing) for the AI model that will act next?''

\begin{figure}[!t]
    \centering
    \includegraphics[width=0.95\columnwidth]{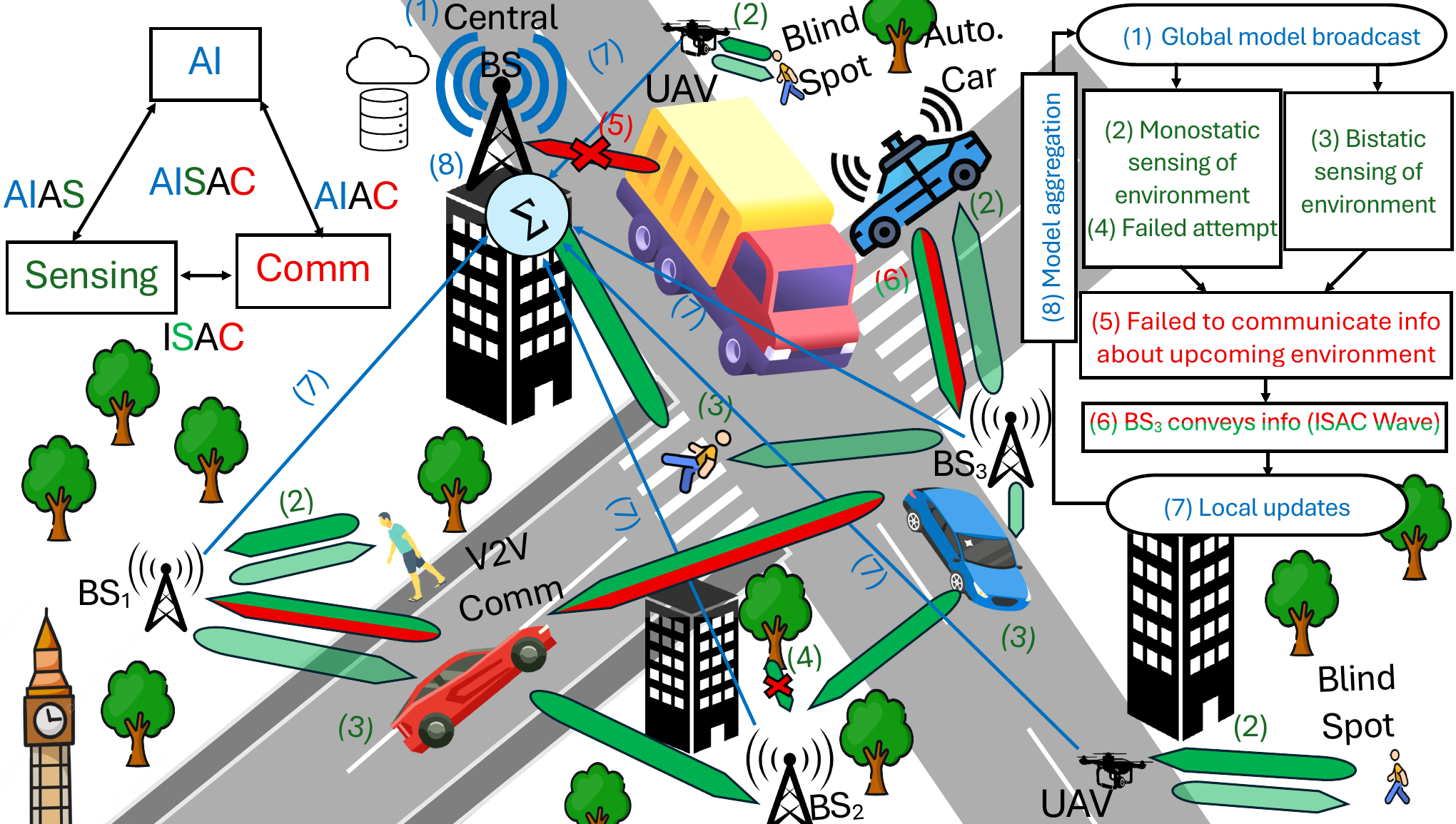}
    \caption{From separate AIAC, AIAS, and ISAC usage scenarios to AISAC, closing the loop between model state, physical-layer acquisition, communication, and future learning.}
    \label{fig:aiac_isac_aisac}
\end{figure}


The rest of this article develops the AISAC viewpoint. We first position AISAC as a closed-loop framework that goes beyond one-way AI-enabled ISAC. The AI model not only optimizes an ISAC system but also changes how future sensing and communication resources are used to acquire and deliver the data and context needed for learning. We then introduce learning alignment as a cross-layer design principle for AISAC, showing why the most accurate sensing configuration in a classical estimation sense is not necessarily the most useful one for training or inference. Finally, we develop a shared-resource AISAC architecture that exposes the three-way tension among sensing-mission accuracy, communication performance, and learning utility, and uses a vehicular edge-intelligence use case to illustrate how closed-loop AISAC differs from sensing-optimal, communication-optimal, and fixed ISAC designs.

\section{From Classical ISAC to AISAC: Landscape and Missing Link}
\label{sec:isac2aisac}
The literature around AI and ISAC is already rich, but its terminology can make the landscape look more unified than it is. We organize it into four categories: \emph{classical ISAC}, \emph{AI for ISAC}, \emph{ISAC-assisted edge intelligence}, and \emph{integrated sensing, communication, and computation (ISCC)}, to clarify the gap that motivates AISAC. 


\emph{Classical ISAC} asks how a wireless system can sense and communicate using shared spectrum, signals, antennas, and hardware. The central tradeoff is usually between a communication metric, such as rate or reliability, and a sensing metric, such as detection probability, localization error, or a Cramer--Rao bound. This line of work established the core dual-function view of wireless signals and clarified why sensing and communication should not be engineered as fully separate radio functions~\cite{LiuJSAC2022,AnLiu_Survey2022_ISAC}.

\emph{AI for ISAC}, also referred to as AI-enhanced ISAC in the literature, uses learning as a design tool for the ISAC system itself. Examples include learning-based waveform design, hybrid beamforming, target classification, clutter suppression, beam prediction, mobility management, multi-agent network control, digital-twin-assisted operation, and autonomous resource management~\cite{VaeziAIISAC, WuAIEnhancedISAC}. This direction is important, especially when the physical models are incomplete or the classical optimization problems are too expensive to solve in real time. However, most AI for ISAC work keeps the AI model outside the data-acquisition loop. In this setting, AI chooses better ISAC policies, but the downstream learning task is not itself the reason why future sensing and communication resources are configured. In other words, AI improves ISAC, but ISAC is not yet designed as a model-dependent acquisition mechanism for improving the next round of learning.

\emph{ISAC-assisted edge intelligence}, also referred to as ISAC for AI in the
literature, reverses the direction. Here, ISAC helps AI by merging sensing and data uploading, reducing the time needed to collect and deliver samples, or providing radio-derived context for edge inference~\cite{ZhangISACEI}. A representative example is the use of ISAC beamforming and time allocation to accelerate edge intelligence by combining sample generation and uploading in one physical process~\cite{LiuISEA2025}. This direction begins to recognize ISAC as a data-generation engine for AI, but it often treats the learning model or task sensitivity as fixed during the ISAC design.

\emph{Integrated sensing, communication, and computation (ISCC)} frameworks add the computing layer explicitly. They jointly allocate sensing, communication, and computation resources to improve learning performance under latency and energy constraints, often through variables such as newly sensed data size, communication power, computation frequency, and over-the-air aggregation parameters~\cite{LiuJSTSP2023,WenISCC2025}. These works are close to AISAC in spirit and are essential references. Their main focus, however, is the joint allocation of system resources for a learning process that is taken as given, without considering all the sensing parameters and their interactions with the evolving model sensitivity. AISAC explicitly models imperfect sensing (multimodal noisy sensors) and adds a sharper feedback question: how should the evolving AI model change future ISAC acquisition itself, especially when sensing errors have a geometry that interacts with the model's loss?

\begin{figure*}[!t]
    \centering
    \includegraphics[width=0.7\linewidth]{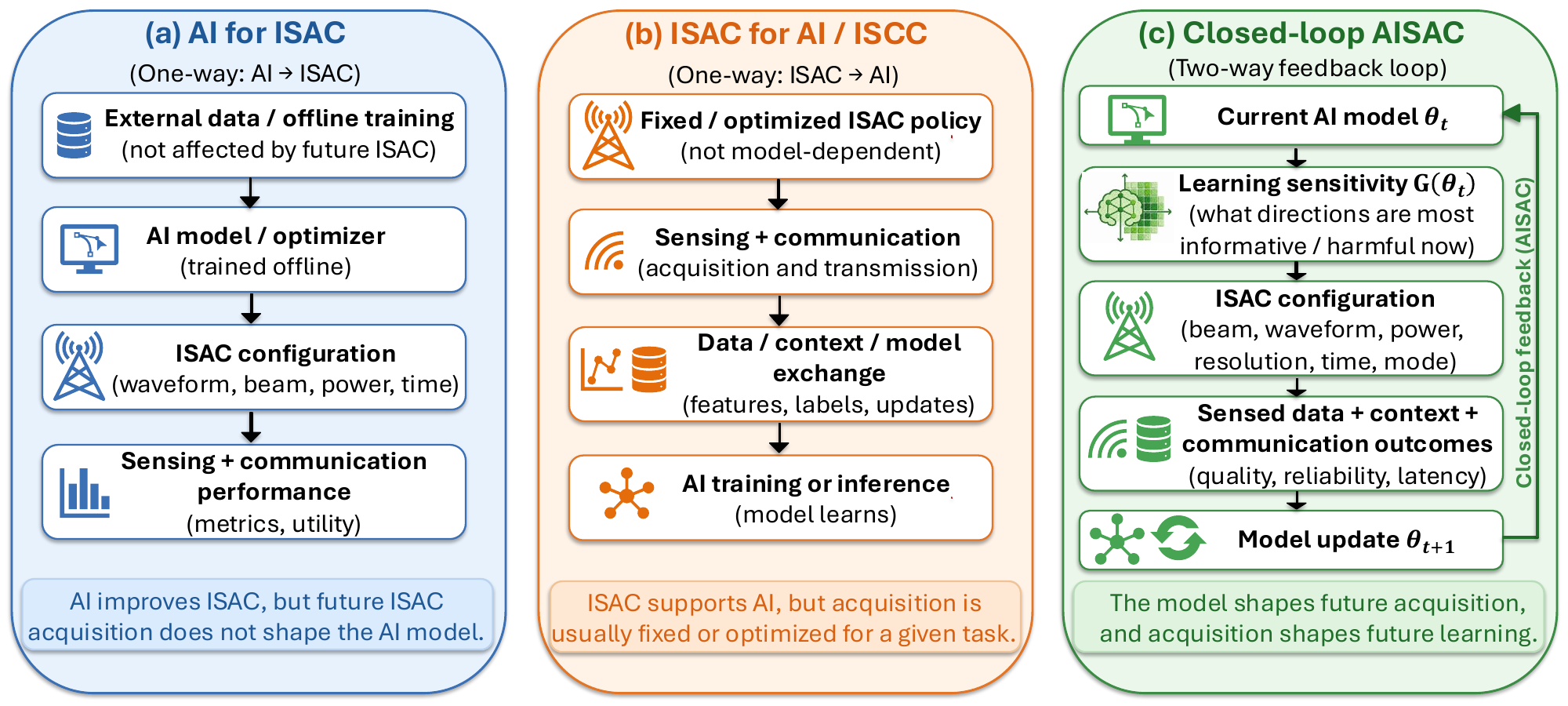}
    \caption{AISAC landscape with closed-loop model state, ISAC acquisition, and future learning.}
    \label{fig:taxonomy}
\end{figure*}

Table~\ref{tab:landscape} and Fig.~\ref{fig:taxonomy} summarize the resulting gap. Prior work often has one of two missing arrows. Either AI optimizes ISAC using externally available data, or ISAC supports AI with a fixed acquisition policy. AISAC closes both arrows: the current model affects the next ISAC configuration, and that configuration changes the sensed data, communication quality, and model evolution that follow.

\begin{table*}[!t]
\centering
\caption{From ISAC to AISAC}
\label{tab:landscape}
\renewcommand{\arraystretch}{1.2}
\begin{tabular}{p{0.10\textwidth}p{0.23\textwidth}p{0.185\textwidth}p{0.20\textwidth}p{0.16\textwidth}}
\toprule
\textbf{Paradigm} & \textbf{Main question} & \textbf{AI role} & \textbf{ISAC role} & \textbf{Closed AI--ISAC loop}\\
\midrule
Classical ISAC & How should sensing and communication share radio resources? & Optional or absent & Dual-function physical layer & No\\
AI for ISAC & How can learning improve ISAC operation? & Optimizer or controller & System being optimized & No\\
ISAC for AI & How can sensing and communication support AI tasks? & Consumer of data and context & Data and connectivity provider & Partial\\
ISCC & How should sensing, communication, and computation be jointly allocated? & Training or inference task & Resource layer for data and updates & Limited or task-specific\\
\textbf{AISAC} & How should AI and ISAC co-adapt over time? & Learner and controller & Data, context, and shared physical resource provider & Yes\\
\bottomrule
\end{tabular}
\end{table*}

\section{AISAC Principle: Why Sensing Distortion Is Not Learning Distortion}
\label{sec:aisac-principles}
AISAC is mainly motivated by a fundamental asymmetry between sensing and communication impairments in learning-driven networks. They enter the learning process at different stages and leave different traces on the model. A communication error typically affects what is exchanged, such as a model update, a feature vector, a gradient, a packet, or a decision. It can often be reduced through coding, retransmission, aggregation design, power control, or scheduling. A sensing error occurs earlier. It changes the observation from which the sample is created. If that sample is stored, reused, or repeatedly involved in local training, the sensing error becomes part of the data distribution seen by the model.

The quantity that ultimately matters in AISAC is not sensing distortion or communication rate alone, but the performance of the AI task that the network exists to serve. We denote this quantity by $\mathcal{E}_{\rm AI}$, which may represent a task loss, inference error, regret, or training metric such as the average gradient norm over the training horizon. AISAC design decisions are therefore evaluated by how they improve $\mathcal{E}_{\rm AI}$, while still satisfying the sensing-mission and communication requirements of the network.

What makes this a system problem is that no single layer controls $\mathcal{E}_{\rm AI}$. At a high level, it can be bounded by the main mechanisms that degrade learning performance,

\begin{align}
\mathcal{E}_{\rm AI}
\lesssim \underbrace{\mathcal{E}_{\rm opt}}_{\text{optimization}} + \underbrace{\mathcal{E}_{\rm sense}}_{\text{data acquisition}} + \underbrace{\mathcal{E}_{\rm comm}}_{\text{model/data exchange}} + \underbrace{\mathcal{E}_{\rm comp}}_{\text{computation}}.
\label{eq:conceptual_bound}
\end{align}
This expression should be read as a high-level bound rather than an exact decomposition or a universal theorem. The terms may interact, and they are not assumed to be independent; rather, they summarize the main mechanisms through which learning over networked sensing systems can degrade. The key point is the separate sensing term. In AISAC, this term is controlled by physical-layer choices that also determine communication performance and sensing-mission accuracy.

A more informative way to think about the sensing term is not only through the total noise power. Suppose an ISAC receiver produces a feature-space observation whose sensing noise has covariance $\bSigma$. To understand how that uncertainty affects learning, we need a companion object in the same feature space. Let $\bG(\btheta)$ be a \emph{gradient-sensitivity matrix}, evaluated at the current model parameters $\btheta$. This matrix measures how strongly perturbations of the acquired features affect the learning objective. Directions associated with large eigenvalues of $\bG(\btheta)$ are those where a small input error produces a large gradient error, and the directions associated with small eigenvalues are those that the model can absorb. Because $\bG(\btheta)$ is evaluated at the current model, it changes as the model trains.

Since $\bSigma$ and $\bG(\btheta)$ live in the same space, they can be compared directly. The sensing harm relevant to learning is then captured by the alignment
\begin{align}
\kappa(\boldsymbol{\theta}) = \trace\big(\mathbf{G}(\boldsymbol{\theta})\boldsymbol{\Sigma}\big). 
\label{eq:kappa} 
\end{align} 
The matrix $\bSigma$ says where the sensor is uncertain. The matrix $\bG(\btheta)$ says where the learning task is \emph{sensitive}. Their trace product measures the overlap between these two geometries. We refer to this quantity as the \emph{learning-alignment penalty}.

Equation~\eqref{eq:kappa} gives the AISAC principle in its simplest form. Two sensing configurations can have the same total distortion, for example, the same $\trace(\boldsymbol{\Sigma})$, while having different values of $\kappa$. The first may inject uncertainty mostly in directions that the current model can tolerate. The second may inject uncertainty precisely in directions that change the gradient, the classifier boundary, or the inferred physical state. Classical sensing metrics may regard the two configurations as similar. AISAC does not.

\begin{figure}[!t]
    \centering
    \includegraphics[width=\columnwidth]{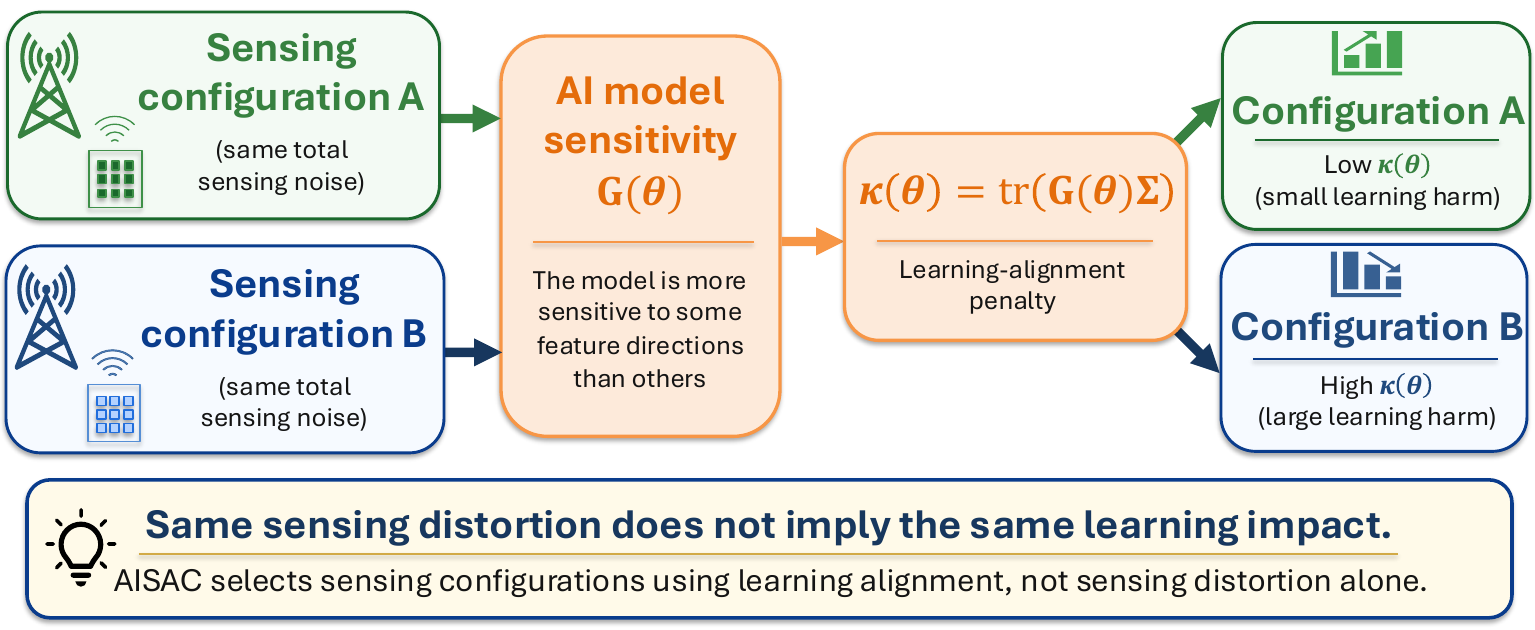}
    \caption{Learning alignment. Two sensing configurations can have equal total noise power but different impact on training or inference because their noise geometries align differently with the model's loss sensitivity.}
    \label{fig:alignment}
\end{figure}

This distinction is especially important in distributed and federated learning. Devices may have different sensors, viewpoints, resolutions, transmit powers, and sampling rates. These differences do not only change the number of samples or the signal-to-noise ratio (SNR). They also change the geometry of the data seen by each local learner. The result is a form of sensing-induced heterogeneity, where the non-independent and non-identically distributed behavior of the training data can come not only from user behavior or class imbalance, but also from the physics of data acquisition. AISAC makes that physics part of the learning system design.

\section{Closed-Loop AISAC: Shared Resources and Three Competing Objectives}
A generic perception-action loop is not enough to define AISAC. Many cyber-physical systems sense the environment, run an algorithm, and act. What separates AISAC is that the algorithm is not a fixed controller acting on perception; it is a model trained from that perception, whose evolving state determines how the next perception is acquired. A second property compounds this. The sensing and communication functions are coupled through the same radio resources. A beam used for environmental sensing also shapes the quality of a communication link. Bandwidth used for range resolution is bandwidth not used for model-update throughput. Transmit power used to illuminate a target is power not available for reliable data exchange. A waveform that is excellent for parameter estimation may not be the one that gives the most useful learning samples. A single configuration decision therefore moves sensing, communication, and learning at once.


These three objectives pull in different directions. The \emph{sensing-mission objective} concerns detection, localization, tracking, classification, or estimation of the physical environment. The \emph{communication objective} concerns rate, latency, reliability, or the quality of model, feature, and control exchange. The \emph{learning objective} concerns the value of the acquired observations and exchanged updates for training, inference, or adaptation. These objectives are generally not aligned.

Let $\rho$ denote a shared ISAC configuration. It may represent a time split, power split, beam allocation, waveform parameter, bandwidth partition, or a more general resource vector. The three objectives can be compactly described as
\begin{align}
\begin{cases}
    J_{\rm sense}(\rho)&= \text{sensing-mission loss}\\
    J_{\rm comm}(\rho)&= \text{communication or model-exchange loss}\\
    J_{\rm learn}(\rho,\boldsymbol{\theta})&= \text{downstream learning loss}
\end{cases}.
\label{eq:three_objectives}
\end{align}
The dependence of $J_{\rm learn}$ on $\btheta$ is central. As the model changes, the learning-sensitive directions change. A resource split or beam that is useful early in training may become less useful later, and a sensor mode that is adequate for one model state may be harmful for another. The learning-optimal ISAC configuration is therefore not fixed; it evolves with the model.

\begin{figure*}[!t]
    \centering
    \includegraphics[width=0.9\linewidth]{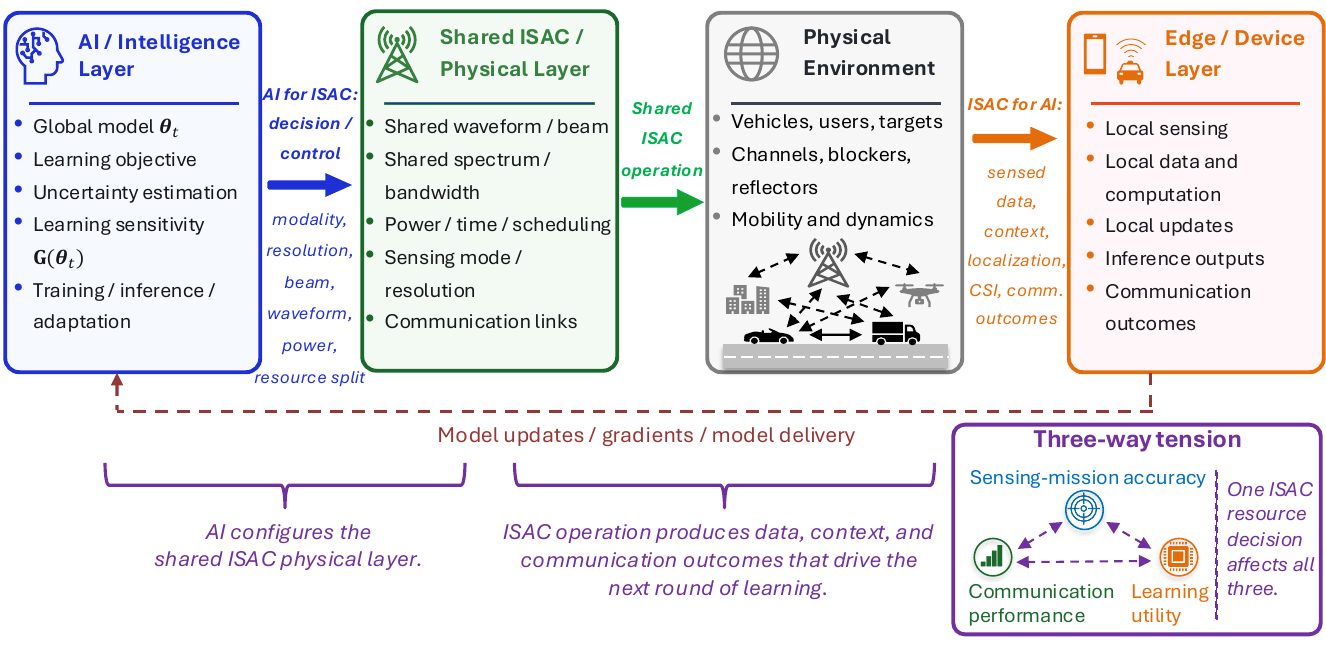}
    \caption{Three-way AISAC tension. The sensing-optimal, communication-optimal, and learning-aligned operating points can be distinct because they value the shared ISAC resource differently.}
    \label{fig:tension}
\end{figure*}

During the training phase, at round~$t$, the current model and its uncertainty determine what the system would benefit from observing next (see the diagram in Fig.~\ref{fig:tension}). The network then chooses an ISAC configuration, including beam, waveform, resource split, sensing mode, sensing resolution, sample count, power level, or scheduling decision. That configuration produces sensed observations and communication outcomes. These are converted into data, context, model updates, or environmental maps, and the AI model is then updated. The new model changes the loss sensitivity and uncertainty, which changes the next ISAC decision.

This loop operates across several timescales. On a fast timescale, the system may adapt beams, waveforms, and detection thresholds for tracking or inference. On a medium timescale, it may schedule users, select samples, allocate sensing and communication resources, or coordinate over-the-air aggregation. On a slow timescale, it may retrain models, update a digital twin, or learn policies for future resource allocation. A practical AISAC design must decide which part of the loop belongs at which timescale. Trying to run every decision at the physical-layer timescale would be unrealistic; running every decision at the model-training timescale would miss the dynamics of the radio environment.

The loop also suggests a small set of metrics that are more concrete than a single weighted utility. Table \ref{tab:metrics} lists four. The \emph{learning-alignment penalty} is the immediate sensing contribution captured by $\kappa$. The \emph{AISAC convergence penalty} is the accumulated cost of sensing, communication, and computation impairments in the learning bound. The \emph{stationarity feasibility margin} asks whether the available sensor hardware, communication resources, and computation budget can reach a target learning accuracy or stationarity level. The \emph{closed-loop adaptation gain} measures what is gained by adapting the ISAC configuration to the evolving model rather than using a fixed or open-loop policy.

\begin{table}[!t]
\centering
\caption{AISAC metrics tied to the closed loop}
\label{tab:metrics}
\renewcommand{\arraystretch}{1.2}
\begin{tabular}{p{0.38\columnwidth}p{0.5\columnwidth}}
\toprule
\textbf{Metric} & \textbf{Role in AISAC}\\
\midrule
Learning-alignment penalty & Measures how strongly sensing uncertainty overlaps with model-sensitive directions.\\
AISAC convergence penalty & Captures how sensing, communication, and computation jointly slow learning or create a performance floor.\\
Stationarity feasibility margin & Indicates whether a target learning condition can be reached under hardware and resource limits.\\
Closed-loop adaptation gain & Measures the benefit of model-dependent ISAC control over fixed or open-loop policies.\\
\bottomrule
\end{tabular}
\end{table}

These metrics are intentionally tied to quantities that can be analyzed or measured, so that the framework specifies not only what an AISAC system should do, but also what it should optimize.

\section{AISAC for Vehicular Edge Intelligence: A Use Case}
\label{sec:case-study}
Vehicular networks offer a natural setting for AISAC because the same infrastructure must sense, communicate, and learn in a changing physical environment. Consider a roadside ISAC base station serving connected vehicles. The base station senses nearby objects, vehicles, pedestrians, and blockages while also communicating with vehicles and other roadside units for control, data exchange, and distributed learning. Vehicles can also sense their surroundings through onboard sensors and exchange information with other vehicles, roadside units, and infrastructure nodes. The learning task may be road-scene classification, blockage prediction, vehicle-type recognition, pedestrian-risk detection, or environment-map refinement. The exact task can vary, but the AISAC structure is the same: sensed observations become learning data, communication links carry model information, and the model affects future sensing and communication decisions.

This use case exposes the three-way tension among sensing, communication, and learning, rather than reducing the problem to a standard federated learning experiment. A useful abstraction is to let the resource variable $\rho_t$ control the ISAC operating mode in round $t$. For example, increasing $\rho_t$ may allocate more time, power, or spatial focus to sensing. This can improve a sensing-mission metric such as tracking accuracy or a Cram\'er--Rao bound, but it may reduce the communication resources available for model-update exchange. More importantly for AISAC, it may change not only the amount of sensing noise but also its feature-space geometry $\bSigma(\rho_t)$. The learning-aligned policy then selects $\rho_t$ based on the current model sensitivity $\bG(\btheta_t)$, not only on the best sensing or communication metric.

A clean evaluation of this setting would compare five policies. A sensing-optimal ISAC would choose the configuration that gives the best sensing-mission accuracy. A communication-optimal ISAC would maximize communication rate or minimize model-update distortion. A fixed compromise policy would use the same resource split over all rounds. An open-loop learning-aware ISAC would use a fixed estimate of learning sensitivity, without updating it as the model evolves. In contrast, closed-loop AISAC updates the ISAC configuration as the model and its sensitivity evolve. A perfect-sensing (no noise in the sensed data) or perfect-link oracle, with error-free channel state information (CSI) and model-update exchange, could also be included as an upper benchmark.



The sensing covariance in this setting should be allowed to change in geometry, not only in magnitude. Otherwise, the learning-alignment principle reduces to an SNR argument. A physically meaningful model can associate narrow beams with accurate but biased coverage of a limited region, wide beams with broader but noisier observations, and different waveform modes with different range, velocity, and feature distortions. In a multimodal extension, radar, camera, LiDAR, acoustic, and radio-frequency features could have distinct covariance structures and sampling costs. The AISAC controller would then decide which sensing mode is most useful for the current model state and the current communication budget.

Distributed learning is especially well matched to this setting. Vehicles and roadside units observe different parts of the environment, generating natural heterogeneity. Over-the-air aggregation or conventional uplink model exchange then consumes the same wireless resources that could be used for sensing. The use case therefore illustrates the full AISAC loop, in which sensing creates the data, communication carries the model information, learning changes the model sensitivity, and the physical layer is reconfigured for the next round.

\section{Open Research Problems and Standards Outlook}
\label{sec:open-problems}
AISAC raises research problems that go beyond existing AI for ISAC, ISAC for AI, and resource-centric ISCC frameworks. Many components of the problem have been studied individually, including AI-based ISAC control, task-oriented sensing and communication, and joint sensing--communication--computation for edge learning. In each of these, the AI model is either an \emph{optimizer} of the ISAC system or a \emph{consumer} of its data. The open challenge is to close the loop, so that the model is both at once; its uncertainty and loss sensitivity influence future sensing and communication acquisition, while the resulting observations, context, and communication outcomes reshape the next model.

\emph{Closed-loop learning theory:} 
Existing convergence analyses for wireless edge learning and ISCC account for communication distortion, computation limits, and resource allocation, and some include data acquisition, typically through a single variable such as the number of newly acquired samples per device or per round~\cite{Liang2025ISCC}. The richer sensing decisions that determine the quality and geometry of the acquired data---modality, sensing resolution, sensing power, waveform, beam, and the resulting noise covariance---are usually not part of the learning-theoretic loop. In AISAC, these decisions become part of the loop. The data distribution depends on the current model because the model influences future sensing and communication acquisition. Model-dependent data distributions have been studied abstractly in decision-dependent and active learning, but their interaction with a physical acquisition layer has not been fully developed, especially when distribution shift is created by beams, waveforms, and power splits and is constrained by a live sensing mission. This setting raises basic questions: when does the loop converge, when can it oscillate, and when can it create blind spots by repeatedly acquiring data about regions where the model is already confident? Online estimation of $\bG(\btheta)$, delayed feedback, heterogeneous sensor geometries, and stability under model-dependent acquisition are the building blocks of this theory.


\emph{Learning-aware physical-layer design:} Existing works at the intersection of AI and ISAC, including AI-based ISAC and task-oriented ISCC, use learning to optimize waveforms, beams, and resource allocation for sensing, communication, and edge-intelligence objectives. AISAC adds a more specific requirement: the physical layer should control the uncertainty that matters to the current model, shaping the sensing covariance so that its alignment with the loss-sensitivity geometry $\bG(\btheta)$ is minimized. This calls for waveform, beamforming, and resource-allocation methods that can expose, estimate, or shape the covariance seen by the AI layer with low enough overhead to support per-round decisions, since both the covariance and $\bG(\btheta)$ evolve as the model trains. The result is a three-way optimization among sensing-mission accuracy, communication performance, and learning utility, in which the sensing mission enters as a hard constraint. The learning-driven acquisition must not compromise sensing functions needed for collision avoidance, localization, or other safety-critical tasks, including public safety applications.


\emph{Distributed and multimodal data acquisition:} In a large network, each device has its own sensor suite, viewpoint, mobility pattern, channel, and computation capability. The resulting heterogeneity is therefore not only statistical, as in conventional distributed learning, but also physical. Radar, camera, LiDAR, acoustic, and radio-frequency features may live in different spaces, be acquired at different costs, and carry different uncertainty geometries. Unlike behavioral non-IID heterogeneity, which is fixed once the data exists, physical heterogeneity is set by acquisition and therefore partly controllable. The network can influence which devices contribute well-aligned data in a given round, and this alignment drifts as devices move and channels change. Exploiting this requires per-device uncertainty and sensitivity summaries that are comparable across modalities, and scheduling that reasons about physical, not just statistical, heterogeneity.


\emph{Foundation-model and semantic AISAC:}
Large multimodal and foundation models may become the intelligence layer that interprets sensed observations, radio context, and network state across many AISAC tasks. This sharpens a question already raised by semantic and goal-oriented communication, but places it inside the AISAC loop: should the shared ISAC physical layer deliver raw measurements, task-specific features, semantic descriptions, or model updates? In AISAC, the answer should depend not only on fixed task semantics, but also on the current model state and its loss sensitivity. A related opportunity is to use generative world models or radio-environment digital twins to predict what a candidate sensing and communication configuration would reveal before spending physical resources. Such models could support model-predictive acquisition, where beams, waveforms, and resource splits are selected to reduce the learning-alignment penalty or expose blind spots. The hard problems are estimating $\bG(\btheta)$ at foundation-model scale, validating semantic sensing quality, and ensuring that large models or autonomous agents do not make unsafe closed-loop physical-layer decisions.

\emph{Privacy, security, and trust:} Beyond the location, motion, human activity, and environmental layout that ISAC sensing already exposes, AISAC introduces a risk specific to the loop. The sensed data stream shapes a model that in turn controls sensing and communication decisions. This creates attack surfaces that neither ISAC nor federated learning exhibits on its own. An adversary could poison the sensing stream so that the learning-aligned resource policy is steered toward the adversary's objective, or exploit the model-to-physical-layer feedback to degrade a sensing mission while appearing to optimize for learning. Defending AISAC therefore means securing the loop as a whole (the data, the model, and the feedback path between them) rather than treating privacy-preserving learning and secure sensing as separable add-ons.

\emph{Benchmarks and reproducibility:} Existing datasets often evaluate sensing, communication, or learning in isolation, and even those that combine modalities are usually static (the data are fixed in advance). Such datasets can support offline evaluation, but they cannot fully capture a closed AISAC loop, where acquisition must react to the model's evolving state. AISAC therefore needs benchmarks in which an evolving physical environment is sensed repeatedly under ISAC configurations that change in response to the learner, capturing resource decisions, sensed observations, communication conditions, model updates, and closed-loop task metrics together. Without such benchmarks, it will be difficult to compare a sensing-optimal ISAC policy and a learning-aligned AISAC policy in a reproducible way.

The standards outlook is timely. 3GPP has studied AI/ML for the NR air interface in TR 38.843 and has also opened standardization-facing work on ISAC for NR in TR 38.765 \cite{3GPP38843,3GPP38765}. These activities show that AI-capable air interfaces and sensing-capable radio networks are both moving toward practical specification. The AISAC question is what must be standardized between them. A closed loop will need model descriptors, uncertainty reports, data-quality metadata, sensing provenance, feedback periodicities, control interfaces, and safety constraints that determine when an AI model is allowed to influence physical-layer acquisition.

\section{Conclusion}
AISAC introduces a fundamental closed-loop design for 6G networks, in which sensing, communication, and learning are optimized as coupled functions rather than separate services. The most useful data for an AI task do not always coincide with the \emph{most accurate} sensing measurements. When sensed observations become training or inference data, what matters is not only how small the sensing error is, but also where it falls. Errors in directions to which the model is sensitive can be costly, while errors in less sensitive directions may have limited effect on the task. This changes the role of waveform, beam, sensing mode, resolution, power, bandwidth, sampling budget, and time allocation. These resources should be configured to satisfy the sensing and communication requirements while aligning the acquired information with the model's current needs. Since those needs evolve as the model trains, AISAC requires a closed loop between model state, physical-layer acquisition, and future learning. AI gives ISAC intelligence, while ISAC gives AI perception, context, and connectivity; AISAC makes this interaction a closed design loop for future wireless networks.

\section{Acknowledgment}
This work was supported in part by the Office of Naval Research (ONR) under grant N00014-21-1-2472 and by the National Science Foundation (NSF) under grants CNS2212565, CNS2225578, and CPS2313109.

\balance

\bibliographystyle{IEEEtran}
\bibliography{IEEEabrv,ref}

@STRING{IEEE_J_STSP       = "{IEEE} J. Sel. Topics Signal Process."}

@STRING{IEEE_J_JSAC       = "{IEEE} J. Select. Areas Commun."}

@STRING{IEEE_J_WCOM       = "{IEEE} Trans. Wireless Commun."}

@STRING{IEEE_M_COM        = "{IEEE} Commun. Mag."}

@techreport{ITU_M2160,
  author      = {{ITU-R}},
  title       = {Framework and Overall Objectives of the Future Development of {IMT} for 2030 and Beyond},
  institution = {International Telecommunication Union},
  type        = {Recommendation ITU-R},
  number      = {M.2160-0},
  month       = nov,
  year        = {2023}
}

@techreport{3GPP38843,
  author      = {{3GPP}},
  title       = {Study on Artificial Intelligence ({AI})/Machine Learning ({ML}) for {NR} Air Interface},
  institution = {3rd Generation Partnership Project},
  type        = {Technical Report},
  number      = {TR 38.843},
  note        = {{Accessed: Sep. 30, 2025}}
}

@techreport{3GPP38765,
  author      = {{3GPP}},
  title       = {Study on Integrated Sensing and Communication ({ISAC}) for {NR}},
  institution = {3rd Generation Partnership Project},
  type        = {Technical Report},
  number      = {TR 38.765},
  version     = {{1.0.0}},
  year        = {2026},
  month       = feb,
  note        = {{Release 20 draft. Accessed: Sep. 30, 2025}}
}

@article{LiuJSAC2022,
  author  = {F. Liu and Y. Cui and C. Masouros and J. Xu and T. X. Han and Y. C. Eldar and S. Buzzi},
  title   = {Integrated Sensing and Communications: Toward Dual-Functional Wireless Networks for {6G} and Beyond},
  journal = IEEE_J_JSAC,
  volume  = {40},
  number  = {6},
  pages   = {1728--1767},
  month   = jun,
  year    = {2022}
}

@ARTICLE{VaeziAIISAC,
  author={Vaezi, Mojtaba and Baduge, Gayan Amarasuriya Aruma and Ollila, Esa and Vorobyov, Sergiy A.},
  journal={IEEE Communications Surveys \& Tutorials}, 
  title={A Tutorial on {AI}-Empowered Integrated Sensing and Communications}, 
  year={2026},
  volume={28},
  number={},
  pages={4980-5013},
  }

@article{LiuJSTSP2023,
  author  = {P. Liu and G. Zhu and S. Wang and W. Jiang and W. Luo and H. V. Poor and S. Cui},
  title   = {Toward Ambient Intelligence: Federated Edge Learning with Task-Oriented Sensing, Computation, and Communication Integration},
  journal = IEEE_J_STSP,
  volume  = {17},
  number  = {1},
  pages   = {158--172},
  month   = jan,
  year    = {2023}
}

@ARTICLE{WenISCC2025,
  author={Wen, Dingzhu and Xie, Sijing and Cao, Xiaowen and Cui, Yuanhao and Xu, Jie and Shi, Yuanming and Cui, Shuguang},
  journal=IEEE_J_WCOM, 
  title={Integrated Sensing, Communication, and Computation for Over-the-Air Federated Edge Learning}, 
  year={2026},
  volume={25},
  number={},
  pages={2748-2762},
  }

@ARTICLE{Brinton-Key6G-CommMag25,
  author={Brinton, Christopher G. and Chiang, Mung and Kim, Kwang Taik and Love, David J. and Beesley, Michael and Repeta, Morris and Roese, John and Beming, Per and Ekudden, Erik and Li, Clara and Wu, Geng and Batra, Nishant and Ghosh, Amitava and Ziegler, Volker and Ji, Tingfang and Prakash, Rajat and Smee, John},
  journal=IEEE_M_COM, 
  title={Key Focus Areas and Enabling Technologies for {6G}}, 
  year={2025},
  volume={63},
  number={3},
  pages={84-91}
}

@ARTICLE{Giordani-6G-ComMag2020,
  author={Giordani, Marco and Polese, Michele and Mezzavilla, Marco and Rangan, Sundeep and Zorzi, Michele},
  journal=IEEE_M_COM, 
  title={Toward {6G} Networks: Use Cases and Technologies}, 
  year={2020},
  volume={58},
  number={3},
  pages={55-61},
  }

@ARTICLE{LiuISEA2025,
  author={Liu, Zhiyan and Chen, Xu and Wu, Hai and Wang, Zhanwei and Chen, Xianhao and Niyato, Dusit and Huang, Kaibin},
  journal={IEEE Communications Surveys \& Tutorials}, 
  title={Integrated Sensing and Edge {AI}: Realizing Intelligent Perception in {6G}}, 
  year={2026},
  volume={28},
  number={},
  pages={2725-2770},
  }

@ARTICLE{WuAIEnhancedISAC,
  author={Wu, Nan and Jiang, Rongkun and Wang, Xinyi and Yang, Lyuxiao and Zhang, Kecheng and Yi, Wenqiang and Nallanathan, Arumugam},
  journal=IEEE_M_COM, 
  title={{AI}-Enhanced Integrated Sensing and Communications: Advancements, Challenges, and Prospects}, 
  year={2024},
  volume={62},
  number={9},
  pages={144-150},
}

@article{Liang2025ISCC,
  author  = {Liang, Yipeng and Chen, Qimei and Zhu, Guangxu and Eldar, Yonina C. and Cui, Shuguang},
  title   = {Communication-and-Energy Efficient Over-the-Air Federated Learning},
  journal = IEEE_J_WCOM,
  year    = {2025},
  volume  = {24},
  number  = {1},
  pages   = {767-782},
}

@techreport{NextGAllianceVerticals2023,
  author       = {{Next G Alliance}},
  title        = {{6G} Roadmap for Vertical Industries},
  institution  = {Alliance for Telecommunications Industry Solutions (ATIS)},
  type         = {White Paper},
  year         = {2023},
  month        = apr,
  note = {[Online]. Available: \url{https://nextgalliance.org/white_papers/6g-roadmap-vertical-industries/}}
}

@article{ZhangISACEI,
  author  = {Zhang, Tong and Li, Guoliang and Wang, Shuai and Zhu, Guangxu and Chen, Gaojie and Wang, Rui},
  title   = {{ISAC}-Accelerated Edge Intelligence: Framework, Optimization, and Analysis},
  journal = {IEEE Trans. Green Commun. Netw.},
  volume  = {7}, 
  number = {1}, 
  pages = {455--468}, 
  year = {2023}
}

@ARTICLE{AnLiu_Survey2022_ISAC,
  author={Liu, An and Huang, Zhe and Li, Min and Wan, Yubo and Li, Wenrui and Han, Tony Xiao and Liu, Chenchen and Du, Rui and Tan, Danny Kai Pin and Lu, Jianmin and Shen, Yuan and Colone, Fabiola and Chetty, Kevin},
  journal={IEEE Communications Surveys \& Tutorials}, 
  title={A Survey on Fundamental Limits of Integrated Sensing and Communication}, 
  year={2022},
  volume={24},
  number={2},
  pages={994-1034},
  }

\end{document}